\documentclass[12pt]{article}

\usepackage{amssymb,amsmath,amsfonts}

\def\proof{\noindent\hspace{2em}{\itshape Proof: }}
\def\QEDclosed{\mbox{\rule[0pt]{1.3ex}{1.3ex}}} 

\def\QED{\QEDclosed} 
\def\endproof{\hspace*{\fill}~\QED\par\endtrivlist\unskip}

\newtheorem{theorem}{Theorem}     
\newtheorem{definition}{Definition}
\newtheorem{proposition}{Proposition}
\newtheorem{lemma}{Lemma}

\newtheorem{remark}{Remark}
\newtheorem{rem}{Remark}

\newcommand{\eqa}{\begin{eqnarray}}
\newcommand{\eeqa}{\end{eqnarray}}
\newcommand{\beq}{\begin{equation}}
\newcommand{\eeq}{\end{equation}}
\newcommand{\nn}{\nonumber}

\def\d{\partial}
\def\n{\noindent}
\def\f{\frac}
\def\H#1{H^{(#1)}}
\def\Hd#1{H^*_{(#1)}}
\def\dpt#1#2{\frac{\partial #1}{\partial t_{#2}}}
\newcommand{\CH}{{{\mathcal{H}}}}
\def\dsl#1{{\displaystyle #1}}
\newcommand{\la}{{{\lambda}}}
\newcommand{\Li}[1]{{{\mathrm{Lie}_{#1}}}}
\def\dpt#1#2{{\frac{\partial #1}{\partial t_{#2}}}}

\begin{document}
\title{Exact Poisson pencils, {\em $\tau$-structures} and topological hierarchies}
\author{Gregorio Falqui, Paolo Lorenzoni\\
\\
{\small Dipartimento di Matematica e Applicazioni}\\
{\small Universit\`a di Milano-Bicocca}
{\small Via Roberto Cozzi 53, I-20125 Milano, Italy}\\
{\small gregorio.falqui@unimib.it, paolo.lorenzoni@unimib.it}}

\date{}

\maketitle

\emph{To Boris Dubrovin in the occasion of his 60th birthday, with friendship and admiration.}\\

\begin{abstract}  We discuss, in the framework of Dubrovin-Zhang's perturbative approach to integrable evolutionary PDEs in 
$1+1$ dimensions, the role of
a special class of Poisson pencils, called exact Poisson pencils. In particular we show that, in the semisimple case, 
exactness of the pencil is equivalent to the constancy of the so-called ``central invariants'' of the theory that were
introduced by  Dubrovin, Liu and Zhang.
\end{abstract}

\section{Introduction}\label{sect0}

Integrable hierarchies of evolutionary PDEs of the form
\begin{equation}\label{fullhi}
q^i_t=V^i_j(q) q^j_x+ \sum_{k=1}^\infty \epsilon^k F_k^i(q, q_x, q_{xx}, \dots, q_{(n)}, \dots)
\end{equation}
have been extensively studied in the last years (see, e.g., \cite{DZ,LZ,DLZ,DLZ2,BPS,L}).

In particular, great attention to  the so-called \emph{topological hierarchies} also because of their relation to the theory of Gromov-Witten invariants, the theory of singularities, and other seemengly unrelated topics of Mathematics and Theoretical Physics.
These hierarchies possess some additional structures: they are bi-Hamiltonian, they admit a tau-structure 
and  satisfy Virasoro constraints \cite{DZ}.
The notion of $\tau$-structure (or $\tau$-function) is perhaps among the oldest ones in the theory of evolutionary equations in $1+1$ dimensions, having 
been introduced by Hirota as the major character in the bilinear formulation of integrable PDEs. Its properties were further exploited by the 
Japanese school (see, e.g.,  \cite{Hir, DJKM,SS}).
In the present approach, the existence of a $\tau$-structure for an integrable hierarchy of $1+1$ evolutionary PDEs
will be understood as the possibility of defining special densities $h^*_i$ for the mutually conserved quantities of the PDEs 
that satisfy the symmetry requirement
\[
\dpt{h^*_i}{j}=\dpt{h^*_j}{i},
\]
where $\dpt{}{k}$ is some suitable one-sequence ordering of the various times of the hierarchy.

Virasoro symmetries are also well known objects of the theory; in particular here we refer to the Virasoro-type algebras 
of additional (explicitly time(s)-dependent) symmetries of the classes of PDEs we are concerned with.
In particular, they gained much attention  in the light of the celebrated results by Kontsevich and Witten \cite{K,W} that identified a 
particular $\tau$-function of the KdV hierarchy with the partition function of 2D Quantum gravity,

As it is well known, the existence of a bi-Hamiltonian structure  means that the equations of the hierarchy can be written in 
Hamiltonian form with respect to two compatible Poisson bivectors $P_1$ and $P_2$
 and that the Poisson pencil $P_2-\lambda P_1$ is 
 a Poisson bivector for any $\lambda$ \cite{M}.
A remarkable result established in \cite{DZ}, and subsequently refined in \cite{BPS} is that,
if the pencil $P_{\lambda}$ is semisimple, (in a sense to be made precise later) and admits a 
$\tau$-function the above requirements fix uniquely 
 the hierarchy once the dispersionless limit
\begin{equation}
q^i_t=V^i_j(q) q^j_x
\end{equation}
and its bi-Hamiltonian structure $(\omega_1,\omega_2)$ are given.
The semisimplicity of the pencil is related to the existence of a special set of coordinates $(u^1,\dots,u^n)$ called
 \emph{canonical coordinates}. If one relaxes the hypothesis of existence of a $tau$-structure, 
 the deformations are parametrized  by certain functional parameters called \emph{central invariants} that  are constants
 in the case of topological hierarchies. 
 In turn, 
 further results in \cite{DLZ}, suggest that the constancy of these central invariants is related with the existence of the $\tau$-function 
 of the hierarchy.\\

In this paper we will show that the Poisson pencil
$$\Pi_{\lambda}=P_2-\lambda P_1$$
of a topological hierarchy is exact, in the sense that there exists a vector field $Z$ (to be called {\em Liouville} vector field of the pencil)
such that
\begin{equation}
\Li{Z} P_2=P_1,\quad\text{ and} \quad 
\Li{Z} P_1=0.
\end{equation}
Moreover, we show that there exists a Miura transformation reducing simultaneously $Z$ to its dispersionless limit:
$$Z\to e=\sum_{i=1}^n\f{\d}{\d u^i}$$
and the pencil $\Pi_{\lambda}$ to the form
$$\omega_{\lambda}+\sum_{k=1}^{\infty}\epsilon^{2k} P^{(2k)}_2.$$
\newline
The hint for our works stems from the observation(s) (to be briefly recalled in Section \ref{sect1}) that the geometry of 
exact bi-Hamitonian manifolds provides {\it somehow for free} the needed "toolkit" requested for the existence of a $\tau$-function 
for the hierarchy.
Indeed, on general grounds, on the one hand the bi-Hamiltonian hierarchies defined on exact bi-Hamitonian manifolds exhibit 
additional symmetries 
of Virasoro type \cite{ZM91}. On the other hand, the action of the Liouville field on
the  Hamiltonian of the hierarchy naturally provides new densities for the conserved quantities.

Actually, we are not going to tackle these problems directly and abstractly as a problem in the general theory 
of Poisson manifolds; rather, we use these "nice" properties of exact Poisson pencils as suggestions for their realization within the
perturbative approach developed in recent years by Boris Dubrovin and his collaborators for the classification problem of $1+1$ evolutionary integrable PDEs of KdV-type. In particular, we borrow from them methods as well as a number of explicit results, with the aim of showing that the geometric notion of exactness of a Poisson pencil can be fruitfully used in this field.  
 
The paper is organized as follows:in Section \ref{sect1} we collect some (more or less known) results about exact Poisson pencils;  then in Section \ref{sect2} we study exact semisimple Poisson pencils of hydrodynamic type and 
we show that for such pencils the vector field $Z$ coincides with the unity vector field $e$ of the underlying Frobenius manifold.
In Sections \ref{sect3} and \ref{sect4} we recall (following \cite{DLZ} and \cite{LZ}) some definitions and results about central invariants and 
bi-Hamiltonian cohomology necessary for the subsequent Section \ref{sect5}
 which is devoted to the
 proof of the main result of the paper. 
 Section \ref{sect6} contains  a brief summary of the paper and some indications of further possible
 steps to generalize the results herewith presented.  
 
\subsection*{Acknowledgments} We warmly thank F. Magri and M. Pedroni for fruitful discussions and useful comments.

\section{Geometry of exact bi-Hamiltonian manifolds}\label{sect1}
In this section we collect some results on the geometry of {\em exact}
bi-Hamiltonian manifolds, and their relations with the hierarchies therein supported. 
It is fair to say that, in one form or the other, these results are known in the 
literature. However, we deem useful to collect them together here, as they somehow provide the 
guiding principle for the arguments contained in the core of the paper.
Let us preliminarily recall a few basic notions.

A {\em bi-Hamiltonian (BH)} manifold\cite{M}  
is a manifold endowed with a pair of compatible Poisson tensors $P_1$, $P_2$ or, equivalently, 
with a pencils of Poisson bivectors $P_\la=P_2-\la P_1$; 
it is well known that this definition entails that separately $P_1$ and $P_2$ are Poisson bivectors, and the the Schouten bracket of $P_1$ and $P_2$ vanishes (this is referred to as the {\em compatibility} condition). 

A sequence of bi-Hamiltonian vector fields $X_i$ satisfying
\begin{equation}\label{eq-rev1}
X_i=P_1 dH_{i+1}= P_2
dH_i,
\end{equation}
with $i$ running in some discrete set of indices, is called a Lenard--Magri sequence. All the vector fields in such a sequence do commute among themselves; equivalently, the functions $H_i$ entering (\ref{eq-rev1}) (the {\em Hamiltonians} of the sequence) are in involution w.r.t. the Poisson brackets defined both by $P_1$ and by $P_2$. 

Following \cite{GZ} we call a Lenard Magri sequence that starts from a Casimir function of one of the Poisson pencil (say, $P_1$) an {\em anchored} sequence; with the term pencil of Gelfan'd--Zakharevich (GZ) type we understand a pencil of Poisson bivectors endowed with $n=\text{dim(Ker} P_1)$ anchored Lenard Magri sequences.   
We remark that  all the Hamiltonians defined by a GZ pencil commute among themselves, even if they belong to {\em different} Lenard Magri sequences.  Also, the pencis of Poisson bivectors entering the Dubrovin-Zhang classification scheme are dispersive deformations of pencils of hydrodynamic type and are all of GZ type. 

We shall herewith consider pencils satisfying an additional geometric requirement.
\begin{definition}\label{exadef}
 Let $P_\la:=P_2-\la P_1$ a pencil of  Poisson bivectors, defined on a BH manifold $\mathcal{M}$. We say that $P_\la$ is 
an {\em exact} Poisson pencil if there exists a vector field $Z\in \mathcal{X}(M)$ such that
\begin{equation}\label{g1}
 P_1=\mathrm{Lie}_Z {P_2};\quad \mathrm{Lie}_Z P_1 (=\mathrm{Lie}_Z^2 {P_2})=0. 
\end{equation}
The vector field $Z$ will be referred to as the {\em Liouville} field   of the exact Poisson pencil.
\end{definition}

We remark that, on general grounds, the Liouville vector field $Z$ is not uniquely defined. For instance adding a bi-Hamiltonian vector field to
 a Liouville vector field one obtains a new Liouville vector field. In the known examples (e.g. in the case of the $A_n$-Drinfel'd-Sokolov hierarchies), 
 there are some natural choices for it. Indeed, in the paper, we shall see that this is the case.
\subsection{Exact BH manifolds and second Hamiltonian function(s)}
Let us consider an exact bi-Hamiltonian manifold, whose Lenard Magri chains be ``anchored'' according to 
the Gel'fand-Zakharevich definition {\cite{GZ}}, that is all chains originate from a Casimir function of $P_1$. 
Let $\CH(\la):=\CH_0+\frac{\CH_1}{\la}+\frac{\CH_2}{\la^2}+\cdots$ a Casimir of the pencil, that is a formal Laurent series 
in $\la$ satisfying
\begin{equation}\label{g2}
P_\la d \CH(\la)=0 (\Rightarrow P_1 d\CH_0=0),
\end{equation}
and consider the pencil of bi-Hamitonian vector fields $X_\la$ of the hierarchy, to be represented as
\begin{equation}\label{g3}
X_\la=P_1 d\CH(\la).
 \end{equation}
\begin{proposition}\label{secham}
 Let $\CH^*(\la):=-\mathrm{Lie}_Z \CH(\la)$; then the one parameter family of vector fields $X_\la$ can be represented as
\begin{equation}\label{g4}
 X_\la=P_\la d\CH^*(\la),
\end{equation}
that is, the deformed Hamiltonians $\CH^*_i=\mathrm{Lie}_Z \CH_i$ define the same GZ foliation of the phase space $\mathcal{M}$.
\end{proposition}
{\bf Proof}. It follows from the straightforward chain of equality
\[ \begin{split}
 0=&\mathrm{Lie}_Z\left(P_\la d \CH(\la)\right)=\mathrm{Lie}_Z(P_\la) d\CH(\la)+P_\la \mathrm{Lie}_Z(d\CH(\la))\\
&=P_1d\CH(\la)+P_\la(d\mathrm{Lie}_Z(\CH(\la))=X_\la-P_\la d\CH^*(\la).\end{split}
\]
\endproof
{\bf Remark}: Exact bi-Hamiltonian pencils, besides having  "historically" provided the first instances of such structures, naturally enter the so-called {\em method of argument translation} related with Lie-Poisson pencils on Lie algebras (see \cite{Man}). 

In the field of evolutionary integrable PDEs, applications of this method  
can be found in \cite{DZ}, \S 3; we notice however that in our case, the Gel'fand-Zakharevich sequences start from Casimir of the "deformed" tensor $P_1$, rather than with Casimirs of the (Lie Poisson) tensor $P_2$. In the case of PDEs, this might be a non-trivial difference.

\subsection{Exact BH manifolds and the Virasoro algebra}
Master symmetries are a very classical topic in the theory of integrable PDEs \cite{Fu,O}.
In \cite{ZM91} it was observed that the Galileian symmetry of the KdV equation could be used as a generator
of a whole (albeit formal) family of  such symmetries, and that such a family is isomorphic to the pronilpotent 
upper subalgebra of the Virasoro algebra, that is, the subalgebra generated by the elements $\ell_k$ with  $k\ge0$.
Here we shall show (see also \cite{AvM94, PvM}) that this is a common feature of all exact bi-Hamiltonian manifolds, and, in particular, 
that the Liouville vector field can be added as the Virasoro generator $\ell_{-1}$.
\begin{definition}\cite{ZM91}\label{confsymdef}
Let $P_\la$ be a Poisson pencil of GZ type, and let $N:=P_2\cdot{P_1}^{-1}$ its formal recursion operator. A vector field
$Y$ is called a conformal symmetry of the pencil if it holds
\begin{equation}\label{cfsy}
\Li{Y} N=N\>.
\end{equation}
\end{definition}
\begin{proposition}
Let $(P_\la, Z)$ be an exact bi-Hamiltonian pencil. Then the field $Y_0:=N\, Z$ is a conformal symmetry of $P_\la$.
\end{proposition}     
\proof
Since $N=P_2\,{P_1}^{-1}$ we have $\Li{Z}(N)=\mathbf{1}$. Now, let us define $Y_0:=N\,(Z)$; 
obviously
\begin{equation}\label{g5}
[Y_0,NX]=\Li{Y_0}(NX)=\Li{Y_0}(N)\, X+N\,\Li{Y_0}(X)=\Li{Y_0}(N)\, X+N\,[Y_0,X].
\end{equation}
The vanishing of the Nijenhuis torsion of $N$ (which, as it is well known to experts in the
theory of Poisson pencil, is implied by the compatibility of $P_2$ and $P_1$) reads, for every pair of vector fields $W,X$
\begin{equation}\label{g6}
[NW,NX]=N\,[NW,X]+N\,[W,NX]-N^2\,[W,X].
\end{equation}
Substituting $Y_0=N\, Z$ in (\ref{g5}) and using the vanishing of the torsion of $N$  we get
\begin{equation}\label{g7}\begin{split}
\Li{N\, Z}(N)&\, X+N\,[N\, Z,X]= N\,[NZ,X]+N\,[Z,NX]-N^2\,[Z,X]=\\
&N\,[NZ,X]+N\, ({\Li{Z}(N)}\, X)+N^2\,[Z,X]-N^2\,[Z,X]\\&\text{which yields }
\Li{N\, Z}(N)\, X=N\, X\> \forall\, X, \quad \text{since } \Li{Z}\, N=\mathbf{1}.\end{split}
\end{equation}
\endproof
As a corollary, we have the following result (see \cite{ZM91} for the full proof, which holds obviously also 
for the slight generalization herewith presented). It is based on the properties
\begin{equation}\label{gx1}
 \Li{Z}N^j=j\,N^{j-1}
\end{equation} 
\begin{proposition}
Let 
\[
Y_j:=N^{j+1}Z (\quad \text{so that } Y_{-1}\equiv Z)
\]
be the family of vector fields obtained formally by the action of the recursion operator on the 
Liouville vector field $Z$. Then the commutation relations
of the Virasoro algebra
\[
[Y_j, Y_k]=(k-j) Y_{k+j}
\]
hold.
\\
\end{proposition}
\subsection{The exact GD pencil and its $\boldsymbol{\tau}$-function}\label{GDn}
The nowadays standard formulation of the n-th ($A_n$) Gel'fand Dickey (henceforth, GD) hierarchy is based on 
its Lax representation (see \cite{Di} 
for a full account of this theory); namely, the phase space is identified with the affine space of differential operators of the form
\[
L=\d^{n+1}+U_n\d^{n-1}+U_{n-1}\d^{n-2}+\cdots+U_1,
\]
that is, the space of monic $n+1$-th order differential operators with {\em vanishing} $n$-th order term.
Its bi-Hamiltonian structure can be represented by means of the Hamilton operators
\begin{equation}\label{nGD}
\begin{split}
&\dot L=P_1(X)=[L,X]_+\\
&\dot L=P_2(X)=(LX)_+L-L(XL)_+-\frac1{n+1}[L, (\d^{-1}[X,L]_{-1})]
\end{split}
\end{equation}
where $X$ represents a one-form on the phase space, that is, a purely non-local pseudodifferential operator. As it is 
customary, the subscript $(\cdot)_+$ refers to the purely differential part of the operator and $(\cdot)_{-1}$ is the 
residue. 
The last term in the second row of (\ref{nGD}) is added to the standard 
Adler-Gel'fand-Dickey Hamiltonian operator in order to preserve the vanishing of coefficient of $\d^n$ of  
Hamiltonian vector fields associated with a generic one-form $X$ (see, e.g., \cite{DLZ} or \cite{BBT} \S 9).

It is well known -- and easily ascertained from  (\ref{nGD})  -- that the Poisson pencil $P_2-\la P_1$ is exact, 
and admits as a  Liouville vector field the field
\[
Z:=\dot{U_1}=1.
\]
It is also well known that the densities of conserved quantities of the  n-th GD hierarchy 
can be collected  in a generating function $h( [U],z)$ of the form
\[
h( [U],z)=z+\sum_{i=1}^{\infty}\frac{h_i( [U])}{z^i}
\]
where $z^{n+1}=\lambda$; 
we use  the symbol $[U]$ as a shorthand notation  for "differential polynomial in the  dependent fields $U_i(x)$";  
the series $h( [U],z)$ is related with the Baker Akhiezer function 
$\psi$ of the theory  by 
\[
\psi=\mathrm{e}^{\int^x h( [U],z) dx}\,\mathrm{e}^{\sum_i t_i z^i},
\]
and is the unique solution of the above form of the Riccati-type equation
\[
h^{(n+1)}+\sum_{j=0}^{n-1}U_{j+1}h^{(j)}{=z^{n+1}},
\]
where $h^{(0)}\equiv 1$ and, by definition, $h^{(k+1)}=\d_x h^{(k)}+h( [U],z)h^{(k)}$.

In  \cite{FMP98} the following representation for the n-th GD (and KP) flows was highlighted:  \\
The GD flows 
imply the
local conservation 
laws
\begin{equation}\label{ourkp}
\dsl{\dpt{}{j}} h( [U],z)=\d_x\H{j}( [U],z),
\end{equation}
where $\H{j}( [U],z)$ are formal series of the form $\H{j}( [U],z)=z^j+\sum_{k=1}^\infty \frac{H_k^j( [U])}{z^k}$
and $z$ is related with the parameter $\la$ of the 
Poisson pencil by $\la=z^{n+1}$
Along the GD flows these "currents" obey the
equations
\begin{equation}\label{CS}
\frac{\partial}{\partial t_j} H^{(k)}( [U],z)=H^{(j+k)}-H^{(j)}H^{(k)}+\sum_{l=
1}^kH^j_lH^{(k-l)}+
\sum_{l=1}^jH^k_lH^{(j-l)}.
\end{equation}
Let us consider the generating function of the densities of the second (or dual) hamiltonian $h^*( [U],z)$.
According to Proposition (\ref{secham}) it must satisfy as well suitable conservation laws, to be written as
\begin{equation}\label{secconslaw}
\dpt{}{j} h^*( [U],z)=\d_x \Hd{j}( [U],z),
\end{equation}
in terms of "dual" currents $\Hd{j}( [U],z)$ that have the form
\begin{equation}\label{Hstar}
 \Hd{j}( [U],z)=jz^{j-1}+\sum_{k=1}^\infty\frac{H^*_{jk}( [U])}{z^{k+1}}.
 \end{equation}
It turns out\footnote{See \cite{CFMP} (where computations are  done in the KP case) for more details.} 
that, if we denote by
${H}_{(l)}=z^{l-1}-\sum_{k\ge 1}H^k_l z^{-(k+1)},$
the dual currents are given by
$\Hd{j}=\sum_{l=1}^j H_{(l)}\H{j-l}.$
By using this representation, and working a bit on the component-wise form of  (\ref{CS}), and in particular on the formula
\begin{equation}
\label{T}
h^*( [U],z)=\frac{\partial}{\partial z}h( [U],z)-\sum_{j=1}^{\infty}\frac1{z^{j+1}}\frac{\partial}{\partial t_j}h( [U],z),
\end{equation}
one can show that
the coefficients $H^*_{jk}$ are symmetric in $j,k$, i.e.,
$H^*_{jk}=H^*_{kj}$, and
along the flows their evolution satisfies
$\displaystyle{\dpt{H^*_{jk}}{l}=\dpt{H^*_{lk}}{j}.}$
Therefore, there exists a function $\tau(t_1,t_2,\dots)$ (independent of the spectral parameter $z$)
such that  
\begin{equation}\label{tau}
H^*_{jk}=\frac{ \d^2}{ \d t_j \d t_k}\log\tau.
\end{equation}
This function is the Hirota $\tau$--function of the GD hierarchy; the outcome that we want to herewith remark is that, in this picture, the $\tau$-function 
appears as the (logarithmic) potential for the densities of conservation laws associated with (\ref{secconslaw}) 
the second Gel'fand-Zakharevich Hamiltonian naturally defined on the exact bi-Hamiltonian phase space of the KdV equation.


\section{The dispersionless case}\label{sect2}
Let us consider an integrable system of the form \eqref{fullhi}, i.e.
\begin{equation}
q^i_t=V^i_j(q) q^j_x+ \sum_{k=1}^\infty \epsilon^k F_k^i(q, q_x, q_{xx}, \dots, q_{(n)}, \dots),
\end{equation}
and consider its
dispersionless (or hydrodynamical) limit. 
The equations of the dispersionless hierarchy have the form
\begin{equation}\label{displim}
q^i_t=V^i_j(q) q^j_x
\end{equation}
For such systems,
 the class of Hamiltonian structures to be considered were introduced by Dubrovin and Novikov. Let us briefly outline the key points
 in their construction. 
Consider functionals 
$${\mathcal F}[q]:=\int_{S^1} f(q^1(x),\dots,q^n(x)) \; dx,\quad \text{ and }\quad
G[q]:=\int_{S^1}g(q^1(x),\dots,q^n(x))\; dx$$
and define a bracket between them as follows:
\begin{eqnarray}\nonumber
\{F, G\}[q]&:=&\iint_{S^1\times S^1}\frac{\delta F}{\delta q^i(x)}\, \omega^{ij}(x,y)\, \frac{\delta G}{\delta q^j(y)}\;dxdy=\\
\label{eq6.eq}
&=&\iint_{S^1\times S^1}\frac{\d f}{\d q^i(x)}\,\omega^{ij}(x,y)\, \frac{\d g}{\d q^j(y)}\;dxdy,
\end{eqnarray}
where $\frac{\delta }{\delta q^i}$ denotes the variational derivative with respect to $q^i$. The bivector $\omega^{ij}(x,y)$ 
has the following (local, hydrodynamical) form 
\begin{equation}\label{eq7.eq}
\omega^{ij}=g^{ij} \delta'(x-y)+\Gamma^{ij}_k q^k_{x} \delta(x-y).
\end{equation}
A deep result geometrically characterizes the conditions for a bracket \eqref{eq6.eq} be Poisson: 
\begin{theorem}\cite{DN84}
If $\det(g^{ij})\neq 0$, then the bracket \eqref{eq6.eq} is Poisson if and only if the metric $g^{ij}$
 is flat and the functions $\Gamma^{ij}_k$ are related to the Christoffel symbols of $g_{ij}$
 (the inverse of $g^{ij}$) by the formula $\Gamma^{ij}_k=-g^{il} \Gamma^j_{lk}$.
\end{theorem}

Let us now consider a pair of Poisson bivectors of hydrodynamic type $\omega^{ij}_1$, $\omega^{ij}_2$,
 associated with a pair of flat metrics $g_1$ and $g_2$.  As shown by Dubrovin in \cite{D07} the flat metrics  define a \emph{bi-Hamiltonian structure
of hydrodynamic type} iff 
\begin{enumerate}
\item the Riemann tensor $R_{\lambda}$ of the pencil $g_{\lambda}:=g^{ij}_2-\lambda g^{ij}_1$ vanishes for any value of $\lambda$;
\item the Christoffel symbols $(\Gamma_{\lambda})_{k}^{ij}$ of the pencil are given by $\Gamma_{(2)k}^{ij}-\lambda\Gamma_{(1)k}^{ij}$.
\end{enumerate}
In this paper we will consider Poisson pencils of hydrodynamic type satisfying
two additional assumptions that can be expressed  on the pencil $g_{\lambda}$ as follows:
\begin{description}
\item[Assumption I:]
The roots
 $u^1(q),\dots,u^n(q)$ of the characteristic equation 
$${\rm det}g_{\lambda}={\rm det}(g_{2}-\lambda g_{1})=0$$ 
are functionally independent. 
\item[Assumption II:]
The Poisson pencil associated to the flat pencil of metrics $g_{\lambda}$ 
according to the Dubrovin-Novikov recipe is an {\em exact} Poisson pencil $\omega_{\lambda}$.
By definition this means that ${\rm Lie}_Z \omega_2=\omega_1$ and  ${\rm Lie}_Z \omega_1=0$
for a suitable vector field $Z$.
\end{description}
The pencil  $g_{\lambda}$ satisfying Assumption I
 is called \emph{semisimple} and the functions $u^i(q)$ are called \emph{canonical coordinates}. 
 It can be shown that, in canonical
 coordinates both metrics are diagonal \cite{Fe}:
$$g_1^{ij}=f^i\delta_{ij},\qquad g_2^{ij}=u^i f^i\delta_{ij}$$ 
 and the the Poisson pencil $\omega_{\lambda}$ becomes
$$\omega_{\lambda}=g_{2}^{ij}(u)\delta'(x-y)+\Gamma^{ij}_{(2)k}u^k_x\delta(x-y)-\lambda\left(g_{1}^{ij}(u)\delta'(x-y)+\Gamma^{ij}_{(1)k}u^k_x\delta(x-y)\right)$$
where the Christoffel symbols vanish if all the indices are different and (assuming $i\ne j$)
\begin{eqnarray*}
&&\Gamma^{ii}_{(1)j}=\f{1}{2}\f{\d f^i}{\d u^j},\,\Gamma^{ij}_{(1)i}=
-\f{1}{2}\f{f^j}{f^i}\f{\d f^i}{\d u^j},\,\Gamma^{ij}_{(1)j}=\f{1}{2}\f{f^i}{f^j}\f{\d f^j}{\d u^i},\,
\Gamma^{ii}_{(1)i}=\f{1}{2}\f{\d f^i}{\d u^i}\\
&&\Gamma^{ii}_{(2)j}=u^i\Gamma^{ii}_{(1)j},\,
\Gamma^{ij}_{(2)i}=u^j\Gamma^{ij}_{(1)i},\,
\Gamma^{ij}_{(2)j}=u^i\Gamma^{ij}_{(1)j},\,
\Gamma^{ii}_{(2)i}=\f{1}{2}f^i+u^i\Gamma^{ii}_{(1)i}.
\end{eqnarray*}

\begin{remark}
In canonical coordinates also the equations of the dispersionless hierarchy become diagonal. 
\end{remark}
The following property will be crucial in the computations we shall perform in the core of the paper

\begin{theorem}\label{th1}
A semisimple bi-Hamiltonian structure of hydrodynamic type is exact if and only if the  condition
\begin{equation}\label{fe}
\sum_{k=1}^n\f{\d f^i}{\d u^k}=0.
\end{equation}
is satisfied.\\ Moreover, in canonical coordinates all the components of the vector field $Z$ are equal to $1$. 
\end{theorem}

\n
\emph{Proof}. By means of a straightforward computation, using formula \cite{LZ}
\begin{eqnarray}\label{Lie}
&&{\rm Lie}_{Z}P^{ij}=\\
&&\sum_{k,s}\left( \partial^s_x Z^k(u(x), \dots) 
\frac{\partial P^{ij}}{\partial u^k_{(s)}(x)}
- \frac{\partial Z^i(u(x), \dots)}{\partial u^k_{(s)}(x)} \partial ^s_x P^{kj}
-\frac{\partial Z^j(u(y), \dots)}{\partial u^k_{(s)}(y)}\partial^s_y P^{ik}\right)\nn,
\end{eqnarray}
we obtain
\begin{eqnarray*}
{\rm Lie}_Z \omega^{ij}_2&=&\left(Z^k\f{\d g_{(2)}^{ij}}{\d u^k}-
\f{\d Z^i}{\d u^k}g_{(2)}^{kj}-
\f{\d Z^j}{\d u^k}g_{(2)}^{ik}\right)\delta'(x-y)+\\
&&\left(Z^k\f{\d\Gamma_{(2)l}^{ij}}{\d u^k}-\f{\d Z^i}{\d u^k}\Gamma_{(2)l}^{kj}-\f{\d Z^j}{\d u^k}\Gamma_{(2)l}^{ik}-g_{(2)}^{ik}\f{\d^2 Z^j}{\d u^k\d u^l}\right)\,u^l_x\,\delta(x-y)=\omega^{ij}_1
\end{eqnarray*}
Similarly we obtain
\begin{eqnarray*}
{\rm Lie}_Z \omega^{ij}_1&=&\left(Z^k\f{\d g_{(1)}^{ij}}{\d u^k}-
\f{\d Z^i}{\d u^k}g_{(1)}^{kj}-
\f{\d Z^j}{\d u^k}g_{(1)}^{ik}\right)\delta'(x-y)+\\
&&\left(Z^k\f{\d\Gamma_{(1)l}^{ij}}{\d u^k}-\f{\d Z^i}{\d u^k}\Gamma_{(1)l}^{kj}-\f{\d Z^j}{\d u^k}\Gamma_{(1)l}^{ik}-g_{(1)}^{ik}\f{\d^2 Z^j}{\d u^k\d u^l}\right)\,u^l_x\,\delta(x-y)=0
\end{eqnarray*}
The vanishing of the coefficients of $\delta'(x-y)$ implies
$${\rm Lie}_Z g_2=g_1,\qquad {\rm Lie}_Z g_1=0,$$
or, more explicitly
\begin{eqnarray*}
&&\left({\rm Lie}_Z g_1\right)^{ii}=Z^k\f{\d f^i}{\d u^k}-2f^i\f{\d Z^i}{\d u^i}=0\\
&&\left({\rm Lie}_Z g_2\right)^{ii}=Z^k u^i\f{\d f^i}{\d u^k}+Z^i f^i-2u^i f^i\f{\d Z^i}{\d u^i}=f^i.
\end{eqnarray*}
Taking into account the first equation,
 the second equation implies 
\beq\label{Lvf}
Z^i=1,\,i=1,\dots,n,
\eeq
and, as a consequence, the first equation reduces to \eqref{fe}. 
It remains to verify
$$Z^k\f{\d\Gamma_{(1)l}^{ij}}{\d u^k}-\f{\d Z^i}{\d u^k}\Gamma_{(1)l}^{kj}-\f{\d Z^j}{\d u^k}\Gamma_{(1)l}^{ik}-g_{(1)}^{ik}\f{\d^2 Z^j}{\d u^k\d u^l}=Z^k\f{\d\Gamma_{(1)l}^{ij}}{\d u^k}=0$$
and 
$$Z^k\f{\d\Gamma_{(2)l}^{ij}}{\d u^k}-\f{\d Z^i}{\d u^k}\Gamma_{(2)l}^{kj}-\f{\d Z^j}{\d u^k}\Gamma_{(2)l}^{ik}-g_{(1)}^{ik}\f{\d^2 Z^j}{\d u^k\d u^l}=Z^k\f{\d\Gamma_{(2)l}^{ij}}{\d u^k}=\Gamma_{(1)l}^{ij}.$$
It is easy to check that both follow from (\ref{fe}). 

%
\endproof

\begin{rem}
In the above computations we have used the same letter ($Z$) to denote a vector field on the manifold $M$ and the corresponding
 vector field on the loop space $\mathcal{L}(M)$. 
\end{rem}

\begin{rem}
The semisimple Poisson pencil of hydrodynamic type associated with a semisimple Frobenius manifold is always exact \cite{D07}. The Liouville vector field in this context is usually denoted by the letter $e$ and called the unity vector field. 
\end{rem}

\subsection{The n--th GD example}
Let us consider the dispersionless limit of the $A_n$ Drinfel'd-Sokolov bi-Hamiltonian structure.
 In this case we have the following generating functions for the contravariant
 components of the metrics of the pencil \cite{SYS,DLZ2}
\begin{eqnarray*}
g_1(q,p) &=&\sum^n_{i,j=1}g^{ij}_{1}p^{i-1}q^{j-1}=\f{\lambda'(p) - \lambda'(q)}{p - q}\\
g_2(q,p) &=&\sum^n_{i,j=1}g^{ij}_{2}p^{i-1}q^{j-1}=
\f{\lambda'(p)\lambda(q)-\lambda'(q)\lambda(p)}{p - q}
+\f{\lambda'(p)\lambda'(q)}{n+1}
\end{eqnarray*}
where
$$\lambda(p)=p^{n+1} + U^{n}p^{n-1} + \cdots+ U^2p + U^1.$$
Clearly, since $\lambda'$ does not depend on $U^1$ and $\frac{\d\lambda}{\d U^1}=1$, 
we have
$${\rm Lie}_Z g_{2}=g_{1},\qquad{\rm Lie}_Z g_{1}=0,$$
with $Z=\frac{\d}{\d U^1}$, that is the Poisson pencil associated with $g_1$ and $g_2$
 is exact. Moreover it is also semisimple. The canonical coordinates $(u^1,\dots,u^n)$
 are the critical values of $\lambda$. If we denote by $v_1,\dots,v_n$ the 
 critical points of $\lambda$ (by definition they do not depend on $U^1$):
$$\lambda'(p)=(n+1)p^{n} + (n-1)U^np^{n-2} + \cdots+ U^2=(n+1)\prod_{k=1}^n(p-v_k)=0,$$
the canonical coordinates are
$$u^i=v_i^{n+1} + U^n v_i^{n-1} + \cdots+ U^2 v_i +U^1.$$
As expected, in canonical coordinates, the vector field  $Z$ reads
$$Z=\sum_{i=1}^n \f{\d u^i}{\d U^1}\f{\d}{\d u^i}=\sum_{i=1}^n \f{\d}{\d u^i}.$$

\section{Central invariants}\label{sect3}
The main problem in the approach of the Dubrovin's school to the theory of integrable systems is the classification of Poisson pencils 
 of the form (see for instance \cite{DZ,LZ,DLZ,DLZ2,L,AL})
\begin{eqnarray*}
&&\Pi^{ij}_{\lambda}=\omega^{ij}_{2}+\sum_{k\ge 1}\epsilon^k\sum_{l=0}^{k+1}A^{ij}_{(2)k,l}(q,q_x,\dots,q_{(l)})\delta^{(k-l+1)}(x-y)\\
&&-\lambda\left(\omega^{ij}_{1}+\sum_{k\ge 1}\epsilon^k\sum_{l=0}^{k+1}A^{ij}_{(1)k,l}(q,q_x,\dots,q_{(l)})\delta^{(k-l+1)}(x-y)\right)
\end{eqnarray*}
where $\omega_1$  and $\omega_2$ are semisimple Poisson bivectors of hydrodynamic type and $A^{ij}_{k,l}$ are differential polynomials of degree $l$. We recall that, by definition,
 ${\rm deg}f(q)=0$ and ${\rm deg}(q_{(l)})=l$.

Two pencils $\Pi_{\lambda}$ and $\tilde{\Pi}_{\lambda}$ are considered equivalent if they are related by a Miura transformation
$$\tilde{q}^i= F_0^i(q)+\sum_{k\ge1}\epsilon^k F^i_k(q,q_x,\dots,q_{(k)}),\qquad{\rm det}\f{\d F_0^i}{\d q^j}\ne 0,\, {\rm deg}F^i_k=k.$$
In the semisimple case \cite{LZ} (that is if $\omega_{\lambda}$ is semisimple) equivalence classes of  equivalent Poisson pencils are labelled by $n$ functional
 parameters called \emph{central invariants}. More precisely two pencils having the same leading order are Miura equivalent if and only if they have
 the same central invariants. In general, the problem
of proving the existence of the Poisson pencil corresponding to a given choice of the leading term $\omega_{\lambda}$ and of the central invariants 
 is still open.

Let us recall the definition of the central invariants of a Poisson pencil. 

At each order in $\epsilon$ the coefficient of the term containing the highest derivative of the delta
 function is a tensor field of type $(2,0)$, symmetric for odd derivatives and skewsymmetric for even derivatives. Consider the formal series
\begin{eqnarray*}
&&\pi^{ij}(p,\lambda,q^1,\dots,q^n)=g_2^{ij}p+\sum_{k\ge 1}A^{ij}_{(2)k,0}p^{k+1}
-\lambda\left(g_1^{ij}p+\sum_{k\ge 1}A^{ij}_{(1)k,0}p^{k+1}\right)
\end{eqnarray*}
and denote by $\lambda^i(q,p)$ the roots  of the equation 
$${\rm det}\,\pi^{ij}(p,\lambda,q^1,\dots,q^n)=0.$$ 
Expanding $\lambda^i(q,p)$ at $p=0$  we obtain  
$$\lambda^i=u^i+\lambda^i_2 p^2+\mathcal{O}(p^4)$$
Following \cite{DLZ2} we can define 
the central invariant $c_i$ as
\beq\label{ceninv}
c_i=\f{1}{3}\f{\lambda^i_2(q)}{f^i(q)}
\eeq
It turns out \cite{LZ,DLZ} that the central invariants $c_i$ depend only on the canonical coordinates $u^i$  and are given by the following expression:
\beq\label{CInv}
c_i(u^i)=\f{1}{3(f^i)^2}\left(Q_2^{ii}-u^i\,Q_1^{ii}+\sum_{k\ne i}\f{(P_2^{ki}-u^i\,P_1^{ki})^2}{f^k(u^k-u^i)}\right),\,\,\,i=1,\dots,n.
\eeq
where $P_1^{ij},\,P_2^{ij},\,Q_1^{ij},\,Q^{ij}_2$ are the components of the tensor fields
$A^{(1)ij}_{2,0},\,A^{(2)ij}_{2,0}$, $A^{(1)ij}_{3,0},\,A^{(2)ij}_{3,0}$ in canonical coordinates. This means that, in such coordinates, the pencil has the following expansion in $\epsilon$:
\begin{eqnarray*}
\Pi_{\lambda}^{ij}&=&\omega_{2}^{ij}+\epsilon\left(P_2^{ij}\delta''(x-y)+\cdots\right)+\epsilon^2\left(Q_2^{ij}\delta'''(x-y)+\cdots\right)+\mathcal{O}(\epsilon^3)\\
&&-\lambda\left[\omega_1^{ij}+\epsilon\left(P_1^{ij}\delta''(x-y)+\cdots\right)+\epsilon^2\left(Q_1^{ij}\delta'''(x-y)+\cdots\right)
+\mathcal{O}(\epsilon^3)\right]
\end{eqnarray*}

As a remark, we notice that we can define central invariants in an alternative way, as 
\beq\label{altdef}
c_i=-\f{1}{3\,f^i}\,{\rm Res}_{\lambda=u^i}{\rm Tr}\,g^{-1}_{\lambda}A_{\lambda}
\eeq
where the tensor $A^{ij}$ is defined by
$$ A^{ij}_{\lambda}=Q^{ij}_{\lambda}+(g_{\lambda}^{-1})_{lk}P_{\lambda}^{li}
P_{\lambda}^{kj}.
$$
with
$$Q_{\lambda}^{ij}=Q_2^{ij}-\lambda Q_1^{ij},\qquad P_{\lambda}^{ij}=P_2^{ij}-\lambda P_1^{ij}.$$
To prove this identity we notice that the identity \eqref{CInv} can be written in terms of the tensor $A^{ij}_{\lambda}$ as 
\beq\label{altdef2}
3c_i(u^i)(f^i)^2=\left\{A^{ii}_{\lambda}\right\}_{\lambda=u^i}={\rm Res}_{\lambda=u^i}\sum_{k=1}^n\f{A^{kk}_{\lambda}}{\lambda-u^k}.
\eeq
and therefore, dividing both sides by $f^i$ and using the properties of residues, we obtain
\begin{eqnarray*}
3c_i(u^i)\,f^i&=&{\rm Res}_{\lambda=u^i}\sum_{k=1}^n\f{A^{kk}_{\lambda}}{f^i(\lambda-u^k)}=\\
&&{\rm Res}_{\lambda=u^i}\sum_{k=1}^n\f{A^{kk}_{\lambda}}{f^k(\lambda-u^k)}=\\
&&-{\rm Res}_{\lambda=u^i}\sum_{k=1}^n (g^{-1}_{\lambda})_{kl}A^{lk}_{\lambda}=
-{\rm Res}_{\lambda=u^i}{\rm Tr}\,g^{-1}_{\lambda}A_{\lambda}.
\end{eqnarray*}

Since the quantity ${\rm Tr}\,g^{-1}_{\lambda}A_{\lambda}$ is a scalar function we can evaluate the components  of the $(1,1)$ tensor field $g^{-1}_{\lambda}A_{\lambda}$
 in an arbitrary coordinate system, compute its trace and then, only at the end of
 the computation, write the result in terms of canonical coordinates. We will use this procedure in the following  examples.
  
\n
{\bf AKNS}. Let us consider the Poisson pencil $\omega_2+\epsilon P_2^{(1)}-\lambda \omega_1$ with
\beq\label{AKNS}
\omega_2+\epsilon P_2^{(1)}-\lambda \omega_1=
\begin{pmatrix}
          (2u\partial_x+u_x)\delta &  v\delta'\\ 
          \partial_x(v\delta) & -2\delta'
         \end{pmatrix}+\epsilon\begin{pmatrix}
          0 &  -\delta''\\ 
\delta'' & 0
         \end{pmatrix}
-\lambda\begin{pmatrix}
          0 & \delta'\\ \delta' & 0 
         \end{pmatrix}
\eeq
where, to compactify the formulas, we write $\delta$ instead of $\delta(x-y)$. 
This is the Poisson pencil of the so-called AKNS (or two-boson) hierarchy.

In this case
$$g_{\lambda}=\begin{pmatrix} 2u&v-\lambda\\v-\lambda
&-2\end{pmatrix}.$$
After some computations we get $A_{\lambda}=\f{g_{\lambda}}{{\rm det}g_{\lambda}}$ 
and therefore, taking into account that
$$u^1=v+\sqrt{-4u},\, u^2=v-\sqrt{-4u}.\qquad f^1=\frac{8}{u_2-u_1},\, f^2=\frac{8}{u_1-u_2},$$
 using formula \eqref{altdef} we obtain 
\begin{eqnarray*}
 c_1&=&-\f{1}{3f^1}{\rm Res}_{\lambda=u^1}{\rm Tr}\,g^{-1}_{\lambda}A_{\lambda}=
-\f{1}{3f^1}{\rm Res}_{\lambda=u^1}\f{2}{{\rm det}g_{\lambda}}=-\f{1}{12}\\
c_2&=&-\f{1}{3f^2}{\rm Res}_{\lambda=u^2}{\rm Tr}\,g^{-1}_{\lambda}A_{\lambda}=
-\f{1}{3f^2}{\rm Res}_{\lambda=u^2}\,\f{2}{{\rm det}g_{\lambda}}=-\f{1}{12}
\end{eqnarray*}

\n
{\bf Two component CH}. Moving $P_2^{(1)}$ from $P_2$ to $P_1$ in the Poisson pencil of the AKNS hierarchy one obtains the following
 Poisson pencil \cite{F,LZ}
\beq\label{PP2CH}
P_{\lambda}
=\begin{pmatrix}
          (2u\partial_x+u_x)\delta &  v\delta'\\ 
          \partial_x(v\delta) & -2\delta'
         \end{pmatrix}-\lambda\begin{pmatrix}
          0 & \delta' -\epsilon\delta''\\ \delta'+\epsilon\delta'' & 0 
         \end{pmatrix}
\eeq
which is the Poisson pencil defining the so called CH$_2$ hierarchy.
The pencil $g_{\lambda}$ and the canonical coordinates are the same of the previous example, while
 $A_{\lambda}=\f{\lambda^2 g_{\lambda}}{{\rm det}g_{\lambda}}$.  Using formula \eqref{altdef} we obtain
\begin{eqnarray*}
 c_1&=&-\f{1}{3f^1}{\rm Res}_{\lambda=u^1}{\rm Tr}\,g^{-1}_{\lambda}A_{\lambda}=
-\f{1}{3f^1}{\rm Res}_{\lambda=u^1}\f{2\lambda^2}{{\rm det}g_{\lambda}}=-\f{(u^1)^2}{12}\\
c_2&=&-\f{1}{3f^2}{\rm Res}_{\lambda=u^2}{\rm Tr}\,g^{-1}_{\lambda}A_{\lambda}=
-\f{1}{3f^2}{\rm Res}_{\lambda=u^2}\,\f{2\lambda^2}{{\rm det}g_{\lambda}}=-\f{(u^2)^2}{12}.
\end{eqnarray*}

\begin{rem}
Notice that in both examples the matrix $g^{-1}_{\lambda}A_{\lambda}$ is the identity matrix times a scalar function. In the first
 case this function is $\f{1}{{\rm det}g_{\lambda}}$ while in the second case it is  $\f{\lambda^2}{{\rm det}g_{\lambda}}$.
\end{rem}

\section{Bi-Hamiltonian cohomology}\label{sect4}
In this section we collect, for the reader's convenience,  some definitions and results about (Bi)-Hamiltonian cohomologies and the Dubrovin-Zhang complex (see \cite{DZ} for full details and proofs). Let $g$ be a flat metric on a manifold $M$ and
 $\omega$ be the associated Poisson bivector of hydrodynamic type. In analogy with
 the case of finite dimensional Poisson manifolds \cite{lichn} one defines 
 Poisson cohomology groups in the following way: 
\begin{equation}\label{eq33bis.eq}
H^j(\mathcal{L}(M), \omega):=\frac{\ker\{d_{\omega}: \Lambda^j_{\text{loc}}\rightarrow \Lambda^{j+1}_{\text{loc}}\}}
{\mathrm{im}\{d_{\omega}: \Lambda^{j-1}_{\text{loc}} \rightarrow \Lambda^j_{\text{loc}}\}}
\end{equation}
where $d_{\omega}:=[\omega,\cdot]$ (the square brackets denote the Schouten brackets) 
and $\Lambda^{j}_{\text{loc}}$ is the space of local $j$-multivectors
on the loop space of the manifold $M$ (see \cite{DZ}
 for more details on the definition of this complex). 
 The space of local multivectors  has a natural decomposition in  components of same degree.
 To determine each component, we recall that, by definition, ${\rm deg}\,\delta(x-y)=1$ and $\d_x$ increases the degrees by one so that 
$${\rm deg}\left(A^{i_1,\dots,i_k}\delta^{(l_2)}(x_1-x_2)\dots\delta^{(l_k)}(x_1-x_k)\right)={\rm deg}A^{i_1,\dots,i_k}+(l_2+\dots
+l_k)+k-1,$$ 
where $A^{i_1,\dots,i_k}=A^{i_1,\dots,i_k}(u(x_1),u_{x_1},\dots)$ is a differential polynomial.
 In this way, for instance, a homogeneous vector field of degree $k$ is a vector field whose components are differential
 polynomials of degree $k$. Since
 the decomposition of $\Lambda^{j}_{\text{loc}}$
 in homogeneous components is preserved by $d_{\omega}$,  we have
\begin{equation}\label{eq34.eq}
H^j(\mathcal{L}(M), \omega)=\oplus_{k} H^j_k(\mathcal{L}(M), \omega).
\end{equation}
For Poisson structures of hydrodynamic type like \eqref{eq7.eq}, it has been proved in \cite{G} (see also \cite{DMS} for an independent proof
 of the cases $n=1,2$) that $H^k(\mathcal{L}(M), \omega)=0$ for $k=1,2,\dots$. 
  The vanishing of these cohomology groups implies that any deformation of a Poisson bivector of a hydrodynamic type  
\begin{equation}\label{DPB}
P^{\epsilon} =\omega +\sum_{n=1}^{\infty}\epsilon^{n}P_{n},
\end{equation}
where $P_k\in \Lambda^2_{k+2, \text{loc}}$ can be obtained from $\omega$ by performing a Miura transformation. 

In order to study deformations of Poisson pencil  of hydrodynamic type  it is necessary to introduce
 bi-Hamiltonian cohomology groups \cite{GZ,DZ,LZ}. For $i\ge 2$ they are defined as
$$H^i_{k}(\mathcal{L}(M),\omega_1,\omega_2)
=\f{{\rm Ker}\left(d_{\omega_1} d_{\omega_2}\,|_{\Lambda^{i-1}_{k,\text{loc}}}\right)}{{\rm Im}
\left(d_{\omega_1}|_{\Lambda^{i-2}_{k-2,\text{loc}}}\right)\oplus{\rm Im}\left(d_{\omega_1}|_{\Lambda^{i-2}_{k-2,\text{loc}}}\right)}.$$  
Liu and Zhang showed that, in the semisimple case,
$$H^2_{k}(\mathcal{L}(M),\omega_1,\omega_2)=0
\quad \forall k\ne 2,$$
and that the elements of
$$H^2_{2}(\mathcal{L}(M),\omega_1,\omega_2)$$
have the form
\begin{equation}\label{qtvf}
d_2\left(\sum_{i=1}^n\int c^i(u^i)u^i_x{\rm log}u^i_x\,dx\right)-d_1\left(\sum_{i=1}^n\int u^i c^i(u^i)u^i_x{\rm log}u^i_x\,dx\right)
\end{equation}
where $c^i(u^i)$ are the central invariants introduced
 in the previous section.  More explicitly, the components of these vector fields,
 in canonical coordinates, are given by
\beq\label{LZVF}
X^i=\sum_{j=1}^n\left[\left(\frac{1}{2}
\delta_{ij}\partial_x f^i + A^{ij}\right)
c^j u^j_x + (2\delta_{ij}f^i - L^{ij})\partial_x(
c^j u^j_x)\right],\,i=1,\dots,n.
\eeq
with
\begin{eqnarray} 
A^{ij}&=&\frac{1}{2}
\left(\frac{f^i}{f^j}\frac{\d f^j}{\d u^i} u^j_x -\frac{f^j}{f^i} \frac{\d f^i}{\d u^j} u^i_x\right)\\
L^{ij}&=&\frac{1}{2}\delta_{ij}f^i +\frac{(u^i - u^j)f^i}{2f^j}
\frac{\d f^j}{\d u^i}.
\end{eqnarray}

We will use these facts later. 

\section{Constant central invariants and exactness}\label{sect5}
 
This section is devoted to the proof of the main result of the paper.
\begin{theorem}\label{mainth}
Let  
\beq\label{Poissonp}
\Pi_{\lambda}=P_2-\lambda P_1=\omega_{2}+\sum_{k=1}^{\infty}\epsilon^{k} P^{(k)}_2-\lambda\left(\omega_{1}
+\sum_{k=1}^{\infty}\epsilon^{k} P^{(k)}_1\right).
\eeq
be a Poisson pencil
whose dispersionless limit $\omega_2-\la \omega_1$ is semisimple and exact. 
Then its central invariants are constant if and only if it is, in the sense of formal series of Poisson pencils, exact.
\end{theorem}
In particular, we recall that Theorem \ref{th1} states that a Poisson pencil of hydrodynamic type is exact if and only if 
the quantities $f^j$ satisfy
\[
 \sum_{k=1}^n \frac{\d f^j}{\d u^k}=0, \quad j=1,\ldots, n.
\]

We split the proof of the main theorem into the proof of some Lemmas.

\begin{lemma}\label{lem0}
Let
$$\Pi_{\lambda}=P_2-\lambda P_1=\omega_{2}+\sum_{k=1}^{\infty}\epsilon^{k} P^{(k)}_2-\lambda\left(\omega_{1}
+\sum_{k=1}^{\infty}\epsilon^{k} P^{(k)}_1\right),\,P^{(k)}_{1,2}\in \Lambda^2_{k+2, \text{loc}}$$
be a Poisson pencil whose dispersionless limit $\omega_{\lambda}=\omega_2-\lambda\omega_1$
 is a semisimple Poisson pencil of hydrodynamic type (not necessarily exact). Let $(c^1,\dots,c^n)$ be the central invariants of $\Pi_{\lambda}$.
 Then there exists a Miura transformation reducing it to the form
\beq\label{pencil}
\Pi_{\lambda}=\omega_{\lambda}+\sum_{k=1}^{\infty}\epsilon^{2k} P^{(2k)}_2,\qquad P^{(2)}_2={\rm Lie}_{X_{(c^1,\dots,c^n)}}\omega_1,
\eeq
with $X_{(c^1,\dots,c^n)}$ given by \eqref{qtvf}. 
\end{lemma}

\n
{\emph Proof}. The lemma is a consequence of the vanishing of the second Poisson 
cohomology group \cite{G,DMS,DZ} associated to Poisson structure of hydrodynamic type
and of the triviality of the odd order deformations \cite{LZ,DLZ}.

Let us restrict our attention to exact Poisson pencils of the form (\ref{pencil}). 
This means that there exists a vector field 
$Z=\sum_{k=0}^\infty\epsilon^{2k} Z_{2k}$ 
(${\rm deg}\,Z_{2k}=2k$) such that
\begin{eqnarray}\label{cond0}
&&{\rm Lie}_Z (\omega_1)=0,\\
\label{cond1}
&&{\rm Lie}_Z (\omega_2+ \sum_{k=1}^\infty\epsilon^{2k}P^{(2k)}_2)=\omega_1.
\end{eqnarray}
From \eqref{cond0} and \eqref{cond1} it follows that
\begin{eqnarray}
\label{exact1}
{\rm Lie}_{Z_0} \omega_1&=&0\\
\label{exact2}
{\rm Lie}_{Z_0} \omega_2&=&\omega_1.
\end{eqnarray}
We have seen (see Theorem \ref{th1}) that this implies \eqref{fe} and that, in canonical coordinates 
 $Z_0^i=1$, that is  $Z_0=e$.

\begin{lemma}\label{lem1}
There exists a Miura transformation preserving $\omega_1$ that reduces $Z$ to $e$.
\end{lemma}

\n
\emph{Proof}. From \eqref{cond0} it follows that
\beq\label{cond4}
{\rm Lie}_{Z_{2k}} (\omega_1)=0,\qquad k=1,2,\dots.
\eeq
This means, in particular, that $Z_2=d_{\omega_1} H_2$ for a suitable functional $H_2$. The Miura transformation generated 
 by the vector field $d_1 \tilde{H}_2$ with
\begin{equation}\label{qwer}
{\rm Lie}_e \tilde{H}_2=H_2\end{equation}
mantains the form of the pencil: $\Pi_{\lambda}\to\tilde{\Pi}_{\lambda}=
\omega_{\lambda}+\sum_{k=1}^{\infty}\epsilon^{2k} \tilde{P}^{(2k)}_2$ 
and reduces $Z$ to the form
\begin{eqnarray*}
Z&=&e+\epsilon^2({\rm Lie}_{d_{\omega_1}\tilde{H}_2}e+d_{\omega_1} H_2)+\mathcal{O}(\epsilon^{4})=\\
&&e+\epsilon^2(d_{\omega_1}(-{\rm Lie}_e \tilde{H}_2)+d_{\omega_1} H_2)+\mathcal{O}(\epsilon^{4})=\\
&&e+\mathcal{O}(\epsilon^{4}).
\end{eqnarray*}
We can apply the same arguments to higher order deformations and construct a Miura transformation
  that maps $Z$ into $e$. 
\endproof

\begin{remark}
For completeness, let us further discuss the  solvability of (\ref{qwer}), that is, of an equation of the form
\beq\label{lieK}
{\rm Lie}_e \tilde{K}=K
\eeq
for the unknown functional $\displaystyle{\tilde{K}=\int_{S^1}\tilde{k}\,dx.}$
In canonical coordinates equation \eqref{lieK} reads
$$\int_{S^1}\sum_{i=1}^n\f{\d\tilde{k}}{\d u^i}\,dx=\int_{S^1}k\,dx.$$
Indeed taking into account the periodic boundary conditions the l.h.s. of \eqref{lieK} is equal to
$$
\sum_{i=1}^n\int_{S^1}e^i\f{\delta\tilde{K}}{\delta u^i}\,dx=\sum_{i=1}^n\int_{S^1}\left[
\f{\d\tilde{k}}{\d u^i}+\d_x\sum_{k=1}^{\infty}(-1)^k\d_x^{k-1}\left(\f{\d\tilde{k}}{\d u^i_{(k)}}\right)\right]\,dx
=\int_{S^1}\sum_{i=1}^n\f{\d\tilde{k}}{\d u^i}\,dx,$$
where $u^i_{(k)}$ is the $k-th$ derivative with respect
 to $x$ of $u^i$. A solution can be found solving the equation
$$\sum_{i=1}^n\f{\d\tilde{k}}{\d u^i}=k$$
for the density of the functional $\tilde{K}$. It is equivalent to the system of equations 
$$\sum_{i=1}^n\f{\d\tilde{A_j}}{\d u^i}=A_j,\qquad\sum_{i=1}^n\f{\d\tilde{B}_{jm}}{\d u^i}=B_{jm},\qquad\dots$$
for the coefficients $\tilde{A}_i,\tilde{B}_{ij},\dots$ of the homogenous differential 
 polynomial 
$$\tilde{k}=\tilde{A}_i u^i_{(N)}+\tilde{B}_{ij} u^i_x u^j_{(N-1)}+\dots$$
 
With a linear change of coordinates  $(u^1,\dots,u^n)\to(w^1,\dots,w^n)$
 we can reduce $\sum_{k=1}^n\f{\d}{\d u^i}$ to $\f{\d}{\d w^1}$. In such coordinates the solution is obtained integrating the coefficients
 of $k$ along $w^1$. Clearly the solution is not unique and in the coordinates $(w^1,\dots,w^n)$
 is defined up to functions of $(w^2,\dots,w^n)$.
\end{remark}

The next lemma shows that the constancy of the central invariants is related to the exactness at the second order of the pencil.

\begin{lemma}\label{lem2}
Let $\Pi_{\lambda}$ be a Poisson pencil of the form \eqref{pencil}. 
Still in the hypotheses of Theorem \ref {th1} (namely, if the condition \eqref{fe} is satisfied), the central invariants of  
$\Pi_{\lambda}$  are constant if and only if the second order condition 
\beq\label{exact3}
{\rm Lie}_{e}P^{(2)}_2=0,
\eeq 
is satisfied.
\end{lemma}

\n
\emph{Proof}. We have the following identity
\beq\label{impid}
{\rm Lie}_{e}P^{(2)}_2={\rm Lie}_{e}{\rm Lie}_{X_{(c_1,\dots,c_n)}}\omega_1={\rm Lie}_{[e,X_{(c_1,\dots,c_n)]}}\omega_1={\rm Lie}_{X_{\left(\f{\d c_1}{\d u^1},\dots,\f{\d c_n}{\d u^n}\right)}}\omega_1
\eeq
 Suppose that ${\rm Lie}_{e}P^{(2)}_2=0$, then, using \eqref{impid}, we have 
$${\rm Lie}_{X_{\left(\f{\d c_1}{\d u^1},\dots,\f{\d c_n}{\d u^n}\right)}}\omega_1=0$$ 
and this implies $\f{\d c_i}{\d u^i}=0,\,\forall i$.  

Suppose now that all the central invariants are constant, then, using
  \eqref{impid} we obtain \eqref{exact3}.

\endproof
\begin{remark}
According to the results of \cite{LZ} and as already stated in Lemma \ref{lem0} we can assume, without loss of generality, that $P^{(2)}_2$
 is given by ${\rm Lie}_X\omega_1$. In this case condition \eqref{exact3} gives the exactness at the second order of the pencil. However in order to prove that the exactness of the pencil
 implies the constancy of the central invariants we have to reduce the Liouville vector field to
 $e$. The reducing Miura transformation, in general, does not preserve $P^{(2)}_2$.
\end{remark}

Lemma \ref{lem2} relates the condition \eqref{exact3} 
to the constancy of the central invariants but does not
 give us any information about the higher order conditions entering the definition of exactness. In order
 to  push our analysis further up in the $\epsilon$ expansion, 
 we need the results  about bi-Hamiltonian cohomology we recalled in the previous section.

\begin{lemma}\label{lem3}
If the condition \eqref{fe} is satisfied, and
the pencil \eqref{pencil} satisfies
$${\rm Lie}_e P^{(2)}_2=0$$
then there exist a Miura transformation such that
$$
\Pi_{\lambda}\to\tilde{\Pi}_{\lambda}
=\omega_{\lambda}+\sum_{k=1}^{\infty}
\epsilon^{2k} \tilde{P}^{(2k)}_2.
$$
with
$${\rm Lie}_e \tilde{P}^{(2k)}_2=0,\qquad k=1,2,\dots$$
\end{lemma}

\n
\emph{Proof}. We construct the Miura transformation by induction. Suppose that the pencil $\Pi_{\lambda}$ satisfies
$${\rm Lie}_e P^{(2k)}_2=0,\dots,N$$
but at the subsequent order, 
$${\rm Lie}_e P^{(2N+2)}_2\neq 0.$$
We show that it is possible to define a Miura transformation such that the transformed 
pencil $\tilde{\Pi}_{\lambda}$ satisfies the above condition, that is, is exact up to order $2N+2$, with Liouville 
vector field still given by $Z=e$.
To construct such a transformation  we will use the following strategy:
\begin{itemize}
\item First we will show that 
$${\rm Lie}_e P^{(2N+2)}_2={\rm Lie}_{X^{(2N+2)}_2}\omega_1$$
and that the vector field $X^{(2N+2)}_2$ belongs to $H^2_{2N+2}(\mathcal{L}(M),\omega_1,\omega_2)$.
 Due to the triviality of this  cohomology group for $N>0$ this implies that
$$X^{(2N+2)}_2=d_{\omega_1} H^{(2N+2)}_2+d_{\omega_2} K^{(2N+2)}_2$$
for two suitable local functionals $H^{(2N+2)}_2$ and $K^{(2N+2)}_2$ having densities which are differential polynomials
 of degree $2N+2$.
\item Second we will show that the pencil $\tilde{\Pi}_{\lambda}$ related to $\Pi_{\lambda}$ by the
 Miura transformation generated by the vector field $d_{\omega_1} \tilde{K}^{(2N+2)}_2$, with 
\begin{equation}\label{qwer2} {\rm Lie}_e \tilde{K}^{(2N+2)}_2=K^{(2N+2)}_2,\end{equation} 
has the required property.
\end{itemize}
Concerning the first point we have to show that
\begin{eqnarray}
\label{d1}
d_{\omega_1}\left({\rm Lie}_e P^{(2N+2)}_2\right)&=&0\\
\label{d2}
d_{\omega_2}\left({\rm Lie}_e P^{(2N+2)}_2\right)&=&0.
\end{eqnarray}
{This can be easily proved using the following consequences of graded Jacobi identity:
\begin{eqnarray}\label{qwer3}
\label{mainId1}
{\rm Lie}_e\,d_{\omega_1}-d_{\omega_1}{\rm Lie}_e&=&0\\
\label{mainId2}
{\rm Lie}_e\,d_{\omega_2}-d_{\omega_2}{\rm Lie}_e&=&d_{\omega_1}. 
\end{eqnarray} 
Indeed, \eqref{d1} follows immediately from \eqref{mainId1} and $d_{\omega_1} P^{(2N+2)}_2=0$. 
To ascertain the validity of  
 \eqref{d2} we first observe that from $[P_2,P_2]=0$ it follows
$$d_{\omega_2} P^{(2N+2)}_2=-\frac{1}{2}\sum_{k=1}^{N}[P_2^{(2k)},P_2^{(2N+2-2k)}];$$
then using \eqref{mainId2} and graded Jacobi we obtain}
\begin{eqnarray*}
d_{\omega_2}\,{\rm Lie}_e P^{(2N+2)}_2&=&{\rm Lie}_e d_{\omega_2} P^{(2N+2)}_2-d_{\omega_1} P^{(2N+2)}_2=\\
&=&-\f{1}{2}\sum_{k=1}^{N}{\rm Lie}_e [P^{(2k)}_2,P^{(2N+2-2k)}_2]=0
\end{eqnarray*}

Concerning the second point (that is, Equation \eqref{qwer2}), we observe that the Miura transformation generated by the vector
 field $\epsilon^{2N+2} d_{\omega_1} \tilde{K}_2^{(2N+2)}$ reduces the pencil to the form 
\begin{eqnarray*}
\tilde{\Pi}_\lambda&=&\omega_\lambda
+\epsilon^2 P^{(2)}_2+\dots+\epsilon^{2N+2}\tilde{P}^{(2N+2)}_2+\mathcal{O}(\epsilon^{2N+4})=\\ 
&&=\omega_\lambda
+\epsilon^2 P^{(2)}_2+\dots+\epsilon^{2N+2}\left(P^{(2N+2)}_2+{\rm Lie}_{d_{\omega_1} \tilde{K}_2^{(2N+2)}}\omega_2\right)+\mathcal{O}(\epsilon^{2N+4})
\end{eqnarray*}
and
\begin{eqnarray*}
{\rm Lie}_e \tilde{P}^{(2N+2)}_2&=&{\rm Lie}_e P^{(2N+2)}_2+{\rm Lie}_e d_{\omega_2} d_{\omega_1}\tilde{K}_2^{(2N+2)}=\cr
&&d_{\omega_1}d_{\omega_2} K_2^{(2N+2)}+d_{\omega_2} d_{\omega_1} {\rm Lie}_e\tilde{K}_2^{(2N+2)}=\cr
&&d_{\omega_1}d_{\omega_2} K_2^{(2N+2)}+d_{\omega_2}d_{\omega_1} K_2^{(2N+2)}=0
\end{eqnarray*}

\endproof

\begin{rem}
{The identity \eqref{mainId1} is the counterpart at the level of the double complex defined by $(d_{\omega_1},d_{\omega_2})$ of the
 exactness of the pencil $\omega_2-\lambda\omega_1$.}  
\end{rem}

\n
Collecting the results of all the previous Lemmas we can finally prove the main theorem.
\newline

\n
{\bf Proof of the main theorem}. Due to lemma \ref{lem0}, without loss generality we can assume that the pencil has the form \eqref{pencil}.  
Suppose that the pencil \eqref{pencil} is exact, i.e.
 it satisfies \eqref{cond0} and \eqref{cond1}. 

Due to lemma \ref{lem1}, performing a Miura transformation preserving $\omega_1$,  
we can reduce $Z$ to $e$. After such a Miura transformation
$$P^{(2)}_2\to{\rm Lie}_{X_{(c_1,\dots,c_n)}}\omega_1+{\rm Lie}_{d_{\omega_1}\tilde{H}_2}\omega_2$$
The exactness of the pencil implies 
$${\rm Lie}_e
\left({\rm Lie}_{X_{(c_1,\dots,c_n)}}
\omega_1+{\rm Lie}_{d_{\omega_1}\tilde{H}_2}
\omega_2\right)=
{\rm Lie}_{X_{\left(\f{\d c_1}{\d u^1},\dots,\f{\d c_n}{\d u^n}\right)}}\omega_1+{\rm Lie}_{d_{\omega_1}({\rm Lie}_e\tilde{H}_2)}\omega_2=0,$$
that is
\beq\label{contrId}
{\rm Lie}_{X_{\left(\f{\d c_1}{\d u^1},\dots,\f{\d c_n}{\d u^n}\right)}}\omega_1=-{\rm Lie}_{d_{\omega_1}({\rm Lie}_e\tilde{H}_2)}\omega_2.
\eeq
The above identity makes sense only if $c^i$=constant   (and hence both sides vanish).  Indeed, \eqref{contrId} tell us
 that the second order deformation 
$$\epsilon^2 {\rm Lie}_{X_{\left(\f{\d c_1}{\d u^1},\dots,\f{\d c_n}{\d u^n}\right)}}\omega_1$$
can be eliminated by the Miura transformation generated by the Hamiltonian vector field
 $\epsilon^2 d_{\omega_1}{\rm Lie}_e\tilde{H}_2$. But, due to
 the results of \cite{LZ}, this is possible only if
 $\f{\d c_i}{\d u^i}=0,\,\forall i$. 
\newline
\newline
Suppose now that the central invariants of the pencil
 \eqref{pencil} are constant.
 Due to lemma \ref{lem2} the pencil satisfies the condition \eqref{exact3}.
 In order to prove that \eqref{pencil} is exact it is enough to prove that it is Miura equivalent to an exact Poisson pencil.  But this follows from lemma \ref{lem3}. 
\endproof
We close this section discussing how the above procedure works for the case of the AKNS hierarchy. 
Let us consider the Poisson pencil \eqref{AKNS}. We have already shown that it has constant central invariants. 
According to theorem \ref{mainth} it is an exact Poisson pencil. The Liouville vector field is $Z=e=\frac{\partial}{\partial v}$.

Notice that
$$\begin{pmatrix}
          0 & -\delta''\\
          \delta'' & 0
         \end{pmatrix}=-{\rm Lie}_{X}\begin{pmatrix}
          (2u\d_x+u_x)\delta & v\delta'\\
          \d_x(v\delta)  & -2\delta'
         \end{pmatrix}$$
where
$$X=\begin{pmatrix}
          0 & \d_x\\ \d_x & 0 
         \end{pmatrix}
\begin{pmatrix}
 \f{\delta H}{\delta\xi}\\
 \f{\delta H}{\delta\eta}
\end{pmatrix},\qquad H=-\int_{S^1}\f{\eta(x)^2}{4}\,dx.$$
This means that the Miura transformation generated by the vector field $X$
 (up to terms of order $\mathcal{O}(\epsilon^3)$)
  reduces the pencil \eqref{AKNS} 
 to the form $P'_{\lambda}=$
\begin{eqnarray*}
&&\begin{pmatrix}
          (2u\d_x+u_x)\delta & v\delta'\\
          \d_x(v\delta)  & -2\delta'
         \end{pmatrix}-\lambda\begin{pmatrix}
          0 & \delta'\\
          \delta' & 0 
         \end{pmatrix}+\\
&&\f{\epsilon^2}{2}{\rm Lie}^2_{X}\begin{pmatrix}
          (2u\d_x+u_x)\delta & v\delta'\\
          \d_x(v\delta)  & -2\delta'
         \end{pmatrix}+\f{\epsilon^3}{6}{\rm Lie}^3_{X}\begin{pmatrix}
          (2u\d_x+u_x)\delta & v\delta'\\
          \d_x(v\delta)  & -2\delta'
         \end{pmatrix}+\dots=\\
&&\begin{pmatrix}
          (2u\d_x+u_x)\delta & v\delta'\\
          \d_x(v\delta)  & -2\delta'
         \end{pmatrix}-\lambda\begin{pmatrix}
          0 & \delta'\\
          \delta' & 0 
         \end{pmatrix}+\f{\epsilon^2}{2}\begin{pmatrix}
          0 & 0\\
          0  & \delta'''
         \end{pmatrix}+
\f{\epsilon^3}{6}\begin{pmatrix}
          0 & -\delta''''\\
          \delta''''  & 0
         \end{pmatrix}
+\dots
\end{eqnarray*}
Notice also that the vector field $Z=e=\f{\d}{\d\eta}$
 is left invariant by the Miura transformation generated by $X$ (indeed $Z$ and $X$ commute). 
 Moreover according to lemma \ref{lem2} ${\rm Lie}_e P^{'(2)}_2=0$.
 
\section{Conclusions and outlook}\label{sect6}
In this paper we elaborated on the circle of ideas connecting exact bi-Hamiltonian pencils, tau structures, and the central invariants of hierarchies 
admitting hydrodynamical limit, as defined by Dubrovin and collaborators. We have provided the characterization of a semisimple
 exact pencil of hydrodynamical type in canonical coordinates. If this is related to a Frobenius manifold, then 
  the Liouville vector field must coincide with the unity vector field. We have shown that  the exactness of the pencil is equivalent to the constancy of the central invariants defined by the dispersive expansion of the Poisson pencil of the hierarchy, and, in particular, that exactness at order $2$ in the $\varepsilon$ expansion is sufficient to ensure exactness at all orders.
We believe that this property is intimately related with the properties of the vector field $e$ that although not belonging to the Dubrovin-Zhang complex, defines an outer derivation of the complex, and satisfies \eqref{qwer3}.

Still, many important examples of bi-Hamiltonian hierarchies of PDEs do not have constant central invariants (and are believed not to admit 
$\tau$-structures, at least in the strong sense herewith understood). 
 Among them the Camassa-Holm equation and its multicomponent generalizations \cite{CH,LZ,CLZ06,F}, and other examples
  belonging to the so called $r$-KdV-CH-hierarchy \cite{MA,AF87,AF88,CLZ09}. 
  In particular in \cite{LZ} it has been shown that the CH equation possesses linear central invariants, while, e.g., the CH$_2$ equation 
  has {\em quadratic} central invariants.
    A natural question would be whether the point of view exposed in the present paper can be applied to characterize these hierarchies.
  Work in this direction is in progress, to be  detailed elsewhere; in particular, according to some preliminary results, 
  this method can be applied almost {\em verbatim} to the case of linear central invariants. It corresponds to the geometric relation, 
  well known in the CH case, 
  \[
  \Li{Z}^2(P_2)=0,\quad\mathrm{ but }\quad  \Li{Z}P_2\neq P_1.
  \]
On the other hand,   in the higher degree case,  the iteration procedure seems to require further condition on the pencil, whose meaning is currently being investigated.


\begin{thebibliography}{10}

\bibitem{AvM94}
M. Adler, P. van Moerbeke, \emph{Compatible Poisson structures and the Virasoro algebra},  Comm. Pure Appl. Math. {\bf 47}  (1994),  no. 1, 5--37. 

\bibitem{AF87}
M. Antonowicz, A.P. Fordy, \emph{Coupled KdV equations with multi-Hamiltonian structures}, Physica D {\bf 28} (1987) 345--357.

\bibitem{AF88}
M. Antonowicz, A.P. Fordy, \emph{Coupled Harry Dym equations with multi-Hamiltonian structures}, J. Phys. A {\bf 21} (1988) 269--275.

\bibitem{AL} A. Arsie, P. Lorenzoni, \emph{On bi-Hamiltonian deformations of exact pencils of hydrodynamic type}, J. Phys. A: Math.
 Theor. {\bf 44} (2011)

\bibitem{BBT} O. Babelon, D. Bernard and M. Talon, \emph{Introduction to classical integrable systems}, Cambridge University Press (2003).

\bibitem{BPS} A. Buryak, H. Posthuma, and S. Shadrin, \emph{A polynomial bracket for Dubrovin-Zhang hierarchies},
 arXiv1009.5351.

\bibitem{CFMP} P. Casati, G. Falqui, F. Magri, M. Pedroni, \emph{The KP theory revisited IV}, 
Preprints SISSA/2  5/96/FM, available online at\\
 {\tt http://www.matapp.unimib.it/~falqui/oldpub/misc.html}.

\bibitem{CH}
R. Camassa and D. Holm, \emph{An integrable shallow water equation with peaked solitons} Phys. Rev. Lett. {\bf 81} 1661-4 (1993).

\bibitem{CLZ06}
M. Chen, S.-Q. Liu, Y. Zhang, \emph{A two-component generalization of the Camassa-Holm equation and its solutions}, Lett. Math. Phys.
 {\bf 75} (2006) 1--15.

\bibitem{CLZ09}
Chen, Ming; Liu, Si-Qi; Zhang, Youjin, \emph{Hamiltonian structures and their reciprocal
 transformations for the $r$-KdV-CH hierarchy},  J. Geom. Phys.  {\bf 59}  (2009),  no. 9, 1227–1243.

\bibitem{Di}
L. A. Dickey, \emph{Soliton equations and Hamiltonian systems}, World Scientific, 1991, Singapore.

\bibitem{DJKM}
E. Date, M. Jimbo, M. Kashiwara, T. Miwa,
{\em Transformation Groups for Soliton Equations.}
Proceedings of R.I.M.S. Symposium on Nonlinear Integrable
Systems--Classical Theory and Quantum Theory
(M. Jimbo, T. Miwa, eds.),
World Scientific, Singapore, 1983, pp.\ 39--119.

\bibitem{DMS} L. Degiovanni, F.Magri, V. Sciacca, \emph{On deformation of Poisson
manifolds of hydrodynamic type}, Comm. Math. Phys. {\bf 253}(1), 1--24 (2005).

\bibitem{DN84} B. Dubrovin, S.P. Novikov, \emph{
On Poisson brackets of hydrodynamic type},
Soviet Math. Dokl. {\bf 279:2} (1984) 294--297.

\bibitem{D07} B. Dubrovin, \emph{Flat pencils of metrics and Frobenius manifolds}, 
In: Proceedings of 1997 Taniguchi Symposium, Integrable Systems and Algebraic
Geometry, Editors M.-H. Saito, Y.Shimizu and K.Ueno, 47-72. World Scientific,
1998.

\bibitem{DZ} B. Dubrovin, Y. Zhang, \emph{Normal forms of integrable
PDEs, Frobenius manifolds and Gromov-Witten invariants},   math.DG/0108160.

\bibitem{DLZ} B. Dubrovin, S.Q. Liu, Y. Zhang, \emph{Hamiltonian peturbations of hyperbolic systems of conservation laws I. Quasi-triviality of bi-Hamiltonian perturbations},
 Comm. Pure Appl. Math. {\bf 59}(4), 559--615 (2006).

\bibitem{DLZ2} B. Dubrovin, S.Q. Liu, Y. Zhang, \emph{Frobenius Manifolds and Central Invariants for
the Drinfeld - Sokolov Bihamiltonian
Structures}, Advances in Mathematics {\bf 219} (3), 780--837 (2008).

\bibitem{D} B. Dubrovin, \emph{Hamiltonian peturbations of hyperbolic systems of conservation laws II},  Comm. Math. Phys. Volume {\bf 267}, Number 1, 117-139, (2006).

\bibitem{F}
G. Falqui, \emph{On a Camassa-Holm type equation with two dependent variables}.  J. Phys. A  {\bf 39}  (2006),  no. 2, 327–342.

\bibitem{FMP98} G. Falqui, F. Magri, M. Pedroni, \emph{Bi-Hamiltonian geometry, Darboux coverings, and linearization of the KP hierarchy},
  Comm. Math. Phys.  {\bf 197}  (1998), no. 2, 303--324.

\bibitem{Fe}
E.V. Ferapontov, \emph{Compatible Poisson brackets
 of hydrodynamic type}, J. Phys. A {\bf 34} (2001) 2377--2388.

\bibitem{Fu}
B. Fuchssteiner, \emph{Mastersymmetries, higher order time-dependent symmetries and
conserved densities of nonlinear evolution equations}, Prog. Theor. Phys. {\bf 70} (1983),
1508--1522

\bibitem{GZ} I. M. Gel'fand and I Zakharevich, \emph{On the local geometry of a Bihamiltonian structure},
 in: Gel'fand Seminar 1990/92, Birkhauser 1993, available 
online at {\tt http://math.berkeley.edu/}$^{\sim}${\tt ilya/papers/bihamiltonian\_1993/res\_web.pdf}

\bibitem{G} E. Getzler, \emph{A Darboux theorem for Hamiltonian operators
in the formal calculus of variations}, Duke Math. J. {\bf 111} (2002) 535--560.

\bibitem{Hir} 
R. Hirota, {\em Exact solution of the Korteweg--de Vries
equation for multiple collisions of solitons}. Phys. Rev. Lett.
{\bf 27} (1972), 1192--1194.

\bibitem{K} 
M. Kontsevich, {\em Intersection theory on the moduli space of curves and the matrix
Airy function}, Comm. Math. Phys. {\bf 147} (1992), 1--23.

\bibitem{lichn} A. Lichnerowicz,\emph{ Les vari\'et\'es de Poisson et leurs alg\`ebres de
Lie associe\'es}, J. Diff. Geom. {\bf 12} (1977) 253--300.

\bibitem{LZ} S.Q. Liu, Y. Zhang, \emph{Deformations of semisimple bihamiltonian structures of
 hydrodynamic type}, J. Geom. Phys. {\bf 54}(4), 427--453 (2005).

\bibitem{L} P. Lorenzoni, \emph{Deformations of bihamiltonian structures
of hydrodynamic type}, J. Geom. Phys. {\bf 44} (2002) 331--375.

\bibitem{M} F. Magri, \emph{A simple construction of integrable systems}, 
J. Math. Phys. {\bf 19} (1978) 1156--1162.
\bibitem{Man}

S.V. Manakov, \emph{Note on the integration of Euler's equations of the dynamics of
an n-dimensional rigid body}, Funct. Anal. Appl. {\bf 4} (1976), 328-329.

\bibitem{MA}
L. Martínez Alonso, \emph{Schrödinger spectral problems with energy-dependent potentials as sources of nonlinear Hamiltonian evolution equations},
J. Math. Phys. {\bf 21} (1980) 2342--2349.

\bibitem{PvM} P. van Moerbeke, \emph{Integrable foundations of string theory}. 
Lectures on integrable systems (Sophia-Antipolis, 1991), 163?267, World Sci. Publ., River Edge, NJ, 1994.

\bibitem{O}
P. Olver, \emph{Applications of Lie groups to differential equations}, Graduate Texts in Mathematics, 107. Springer-Verlag, New York, 1986

\bibitem{SYS}
K. Saito, T. Yano, J. Sekeguchi, \emph{On a certain generator system of the ring of invariants of a finite reflection group},
Comm. Algebra {\bf 8} (1980) 373--408.

\bibitem{SS}
M. Sato, Y. Sato,
{\em Soliton equations as dynamical systems on infinite--dimensional 
Grassmann manifold.}
Nonlinear PDEs in Applied Sciences (US-Japan Seminar, Tokyo), 
P. Lax and H. Fujita eds., 
North-Holland, Amsterdam, 1982, pp.\ 259--271.

\bibitem{W}
E.Witten, \emph{Two-dimensional gravity and intersection theory on moduli space},
Surv. in Diff. geom. {\bf 1} (1991), 243--310.

\bibitem{ZM91} 
J.P. Zubelli, F. Magri \emph{Differential equations in 
the spectral parameter, Darboux transformations and a hierarchy of master symmetries for KdV},  
Comm. Math. Phys.  {\bf 141}  (1991),  no. 2, 329--351. 

\end{thebibliography}
\end{document}